\documentclass[oneside,11pt]{report}
\usepackage{epsf}
\begin{document}
\title{
Angular and three--dimensional correlation functions, determined from the Muenster Red Sky Survey}
\author { Peter Bosch\'an\\
Department of Theoretical Physics, Westf\"alische Wilhelms Universit\"at, \\
D48149 Muenster, Germany\\
boschan@uni-muenster.de} 
\maketitle
\begin{abstract}
We measured the two dimensional galaxy--galaxy correlation function from the
Muenster Red Sky Survey (Ungruhe 1999). This survey has a slightly larger 
surface as the APM survey and complete up to r$_F<$18.3 mag 
($z_{med} \sim $0.14). The large dynamical
range of the survey made it possible to examine sub-catalogues with different
limiting magnitudes. The positive part of our measured angular correlation 
functions (after the colour correction) agree excellently with those measured 
by Maddox, Efstathiou and Shutherland 1996 from the APM galaxy catalogue 
(Maddox et al. 1990). From the measured angular correlation functions the 
three dimensional two-point correlation function $\xi(r)$ was calculated. 
We have chosen the evolution parameter 
$\epsilon$ so that the change of the correlation length $r_0\,$ with the depth
of the sub-catalogue were minimal. This prescription gives 
$r_0$ = 5.82$\pm$0.05 h$^{-1}$ Mpc and a very high, positive evolution parameter
$\epsilon$ = 2.6. The position of the point, where the 3d correlation function
breaks away from the power law depends on the depth of the sub-catalogues: 
for low magnitude limits the break-down comes earlier, for higher later.  
\end{abstract}
\normalsize
1. Introduction\\[2mm]
Since almost half a century the two-point correlation 
function has been the most popular statistic to describe 
the galaxy clustering in astrophysics. It has been firmly established
(Peebles 1980 and references therein) that the three dimensional two-point 
correlation function (3d CF) is of the form  $\xi(r) = (r_0/r)^{\gamma}$ in
a wide range of distances, breaks away from the power law form at certain
distance $r_1$ and becomes negative at $r_2$.
The 3d CF can be measured by using magnitude or volume 
limited three dimensional galaxy catalogues or can be calculated from the
2d CF, measured from large surface two dimensional catalogues. 
The parameters $r_0\,$ and $\gamma$ measured from local and medium deep 
catalogues are $r_0 \approx 5 h^{-1}$ Mpc $\gamma = 1.7-1.8$. Some authors 
estimate also the parameter $r_2$ . Baugh 1996 finds by using the APM galaxy
survey (Maddox et al. 1990) $r_2 \approx 30 h^{-1}$ Mpc, Tucker et al. 
1997 measure from the LCRS (Shechtman et al. 1996) $r_2\sim 30 - 40 h^{-1}$ Mpc
while Cappi et al. 1998 on the basis of the Southern Sky Redshift Survey 
(SSRS2, da Costa et al. 1994) find that the  first zero point of the 
autocorrelation function $\xi(r)\,$ increases linearly with the sample depth.
Several authors notices that the 3d CF rises above the power law before
it breaks down (Baugh 1996, Guzzo et al. 1991).

To measure the detailed form of the $r_0(z)\,$function 
large surface deep 3d catalogues are needed. At present the deep 3d surveys 
(say $z_{med} >$  0.3)  contain several hundred galaxies and are not large 
enough for this purposes. Usually the evolution of the CF is parametrised 
and the parameter is measured. The customary parametrisation of the correlation
length $r_0$ as a function of the redshift is
\begin{equation}
   r_0(z) = r_0 (1+z)^{-3-\epsilon \over \gamma}.       \label{eq:r0z}
\end{equation}
This relation was proposed in 1977 (Groth \& Peebles 1977) when the
observable redshifts were much smaller than 1.
For $z \ll 1\,$the first term of expansion  of $r_0(z)$ by $z$ is
$r_0(z) = r_{0}(0) (1 -{3 + \epsilon \over \gamma} z)\,$ ($r_0\,$ measured
in physical coordinates). However, for large $z$ with 
$\epsilon=$ const. (\ref{eq:r0z}) has no solid physical basis. 
In a model proposed by Ma 1999  $\epsilon\,$ varies from $\gamma - 1$ at high
redshift trough 0 at medium $z\,$ down to  $\gamma$ - 3 for low redshift. 
The first value corresponds to the linear grows of the perturbations (over
the general expansion), $\epsilon = 0\,$ if the structures are fix in 
comoving coordinates and $\epsilon =  \gamma$ - 3 when the clustering
is stabile in physical coordinates. The canonical value of the correlation
length is $r_0(0) = 5.4\, h^{-1}$Mpc (Davis \& Peebles 1983). From small 
catalogues in general, one can only measure only a combination of $r_0\,$ 
and $\epsilon\,$ (Le F\'evre et al. 1996, Carlberg et al. 1997, 
Connolly et al 1998, Small et al. 1999)  therefore it is difficult to 
compare the results. The $\epsilon\,$ evolution parameter can be roughly 
estimated from the equation 
$r_0(z_{med}) =r_0(0) (1 + z_{med})^{-{3 + \epsilon \over \gamma}}$ , 
where $z_{med} \,$  is the median
redshift of the catalogue, $r_0\,$ measured in physical coordinates.
It is clear that the amplitude of the 3d CF decreases strongly with 
increasing redshift and for fix $r_0(0)\,$ the measured $\epsilon\,$ 
is between 0 and 2.6.
Postman et al. 1998 determined the 3d CF from an I selected 2d catalogue.
From the I$\leq$20 sub-catalogue they found $r_0 = 5.2 \pm 0.4 h^{-1}$ Mpc
and $\epsilon\,$ positive. Deeper the catalogue (I$\leq$23, $z_{med}$ = 0.5),
$\epsilon\,$ is smaller ie. the evolution is slower what is in contradiction
with the results of Ma 1999. Small et al. 1999 find no evidence that the
comoving correlation length changes between  the redshift intervals 
(0.2 - 0.3) and (0.3 - 0.5).   

From a large surface catalogue 
with large dynamical range (as APM or MRSS) one can separate a number of 
sub-catalogues eg. different limiting magnitudes $m$. In this case by taking
the same $r_0$ for every  sub-catalogue, one can determine the parameter 
$\epsilon$ (at least for $z \ll$1).

A further important question is, where the CF breaks away from scaling, 
at which distance $r$ becomes negative and are there measurable 
anti-correlations and correlations after that point. These problems are 
related with the  question of the size of the largest structures in the 
Universe. Wu, Lahav \& Rees 1998 and 1999 find using several indicators that
on scales larger than 300 h$^{-1}$ Mpc the fluctuations become negligible.

Pietronero and coworkers (Coleman \& Pietronero 1992, Sylos--Labini et al.
1998, Joyce et al. 2000 and references therein) question the applicability 
of the CF to characterise the large scale structure of the Universe. They 
state, on the basis of conditional probability analysis of different catalogues
that the correlation length $r_0\,$ for clustering of galaxies do not 
characterises intrinsic properties of the distribution, but the particular 
sample one is considering and  no statistical evidence for the usually
assumed homogeneity on a scale large enough.

In this paper we determine the two-point correlation function from the Muenster
Red Sky Survey (MRSS, Ungruhe 1999). Section 2 contains a brief description
of the survey. Section 3 presents the measuring methods of the 2d CF, the 
analysis of the results and comparison with the APM 2d correlation function as
determined by Maddox, Efstathiou and Shutherland 1996 ( in the following MES).
In section 4 we calculate the 3d CF from different sub-catalogues using the 
Limber-equation, discuss the scaling properties of the CFs and determine
the correlation length $r_0$, as well as the evolution parameter $\epsilon$. 
The interpretation of the results is discussed in Section 5.



The two-point  CFs give only partial information about the galaxy distribution,
for more detailed description higher order CFs are needed. On this correlation
function we report elsewhere.\\[2mm]

2. The catalogue\\[2mm]

The Muenster red sky survey (MRSS, R. Ungruhe 1999) is based on 217 
contiguous plate
of the ESO  Southern Sky Atlas R, covering more than 5000 square degree.
The plates were digitalised using two PDS 2020 GM$^{plus}$  machine of the 
Astronomical Institute Muenster. The image surface brightness profiles
were used to distinguish galaxies from stars and other objects.
The catalogue  contains more than 7 million galaxies and out of these
about 2.3 million brighter than r$_F$=19.5 mag, but to avoid the "last
magnitude effect" we do not use in the determination of the CF the galaxies
fainter then 18.5 mag. After 
corrections the magnitudes are accurate to $\le \,$ 0.05 mag across
each of the 217 plate. Images in the overlapping regions of neighbouring plates
are used to establish a uniform magnitude system over the entire survey.
In order to calibrate the matched magnitudes, CCD photometrie of 1766 galaxies
and 1876 stars, homogeneously distributed on 92 plates were used. The 
estimated plate to plate zero-point-error is less than 0.10 mag. The 
extinction was corrected using the maps of  Schlegel et al. 1998. The 
catalogue is complete up to 18.3 mag (Ungruhe 1999). 

The catalogue is complemented with galaxy images digitalised in pixels
with 1.01 arcsec edge length. The pictures of objects were stored
in frames (out-cuts) with side length  21, 51 and 101 pixels, choosed 
by the software. The computer classification of about 2.6 million 
objects were carefully controlled visually (Ungruhe 1999), what reduce 
essentially the possible star--galaxy misclassification. \\
The corresponding J-atlas was digitalised at two UK institutions:
at the Institute for Astronomy in Cambridge using the APM machine 
(186 Schmidt-plate, Maddox et al. 1990) and at the Royal Observatory Edinburgh
(154 Schmidt-plate) with the COSMOS machine  Stevenson et al. 1985. 
Both catalogues are included in MRSS that makes possible the comparison of
the positions and other quantities, as well as the determination of the 
r$_F$ - b$_J$ colours of the galaxies. Because of the large surface of the 
survey the integral corrections are negligible.

The F band has some advantages over the J band: the vignetting
and desensitization corrections are 2 - 3 times smaller, consequently
the magnitudes towards the edges of plates are better determined. 
Furthermore, the K - correction is also smaller and  varies mildly with the 
depth of the catalogue. In spectral range of the MRSS
the average  r$_F$ - b$_J$ is 1.5 mag (Jones et al. 1991).
In addition the MRSS catalogue contains other data such as surface brightness, 
ellipticity, length of small and large axes, large-scale alignments etc.
Schuecker et al. (1990, 1994) measured the objective prism 
redshifts for 900 000 galaxies, and  Spiekermann 1994 gave an automated 
morphology classification of all galaxies brighter than r$_F$ = 18.5 mag.\\[2mm]

3. Measuring the correlation function\\[2mm]

As it was emphasised by Maddox, Efstathiou and Shutherland (1996), 
large areas of the sky are required to determine the 2d CF $w$ reliably 
on angular scales of a few degrees. They used  80 to 120 contiguous plates to 
compute the CF.  Here we determine the $w(\theta;m)\,$ from 152 contiguous
plates.

For measuring the galaxy - galaxy correlation function  we use 
two essentially different, independent methods: (i) count in  annuli around 
galaxies and (ii) count in cells.\\[2mm]

3.1 Count in annuli around galaxies\\[2mm]

The most direct method to measure the over-density around galaxies is to
count the number of galaxies between circles of radius $\theta_i$ and 
$\theta_{i+1}$. The corresponding density can be calculated by dividing
by the surface of the annulus. If the first galaxy is near to
the border of the sample only a part of the annulus is taken
into account being inside the sample.  For very small angular distances 
this method is more precise than measuring the surface by the number
of random points in the annulus. This method does not need gridding
of the data, it is simple and fast to apply but,  as it was shown
(Hamilton 1993, Maddox et al. 1996), subject of first order errors in
the galaxy density contrast. We use this estimation only for  small angular 
scales $\theta < 1^o$, when the amplitude is high. \\[2mm]

3.2 Count in annuli around bins\\[2mm]

For larger angular
distances we do not need the exact position of the galaxies.
We binned the galaxies in square cells in an equal area projection, and
applied the count in annulus method for the cells. A cell belongs to
a given ring when its centre is inside the ring. The surface of an annulus 
is taken to be equal to the  the sum of the surfaces of cells in the annulus.
That is we use the usual estimator
\begin{equation}
	w(\theta) = {<n_i n_j> \over <n_i><n_j>} - 1.    \label{eq:cf}
\end{equation}
 In the angular range 0.08$^o \, \theta < \,0.8^o\,$ we use a grid of
edge length $s = 0.01^o$ , while for $0.5^o < \theta \le 4^o \,$ we have
$ s = 0.1^o\,$ and 
$s =  0.25^o$ for $ \theta > 4^o$. The overlapping parts  of the CFs 
calculated in different intervals  agree well with each others.\\[2mm]

3.3 The count-in-cells method\\[2mm]

Another possibility to determine the CF is to measure its average on square
cells of edge $\theta$:
\begin{equation}
   \overline {w}(\theta) = {1 \over {\theta}^4} \int^{\theta}_0 d^2 \theta_1
   \int^{\theta}_0 d^2 \theta_2 w(|\bf \theta_1 - \theta_2|).\label{eq:avw}
\end{equation}
This quantity can be expressed in terms of the moments of galaxy-number 
in a cell as follows  (Rubin 1957, Peebles 1980)
\begin{equation}
	\overline w(\theta) = {\overline {N^2}- \overline{N}^2- 
	\overline {N} \over \overline {N}^2}.   \label{eq:avwavN}
\end{equation}
\vspace{2mm}
To compute $\overline {N} = \overline {N(\theta)}\,$ we put a grid with
edge length of $\theta \,$ on the equal-area-projection. In order to be sure 
not to cut structures in parts we shift the grid in steps $\ll \theta $ in 
$\delta$ and $\alpha$ directions and count the galaxies in cells for every 
position. By calculating the averages the catalogue is
heavily oversampled. The -- mostly beneficial -- consequences of oversampling 
was discussed by Szapudi et al. 1997.\\[2mm]
From the average of the CF one can determine the true CF by inverting the 
integral (\ref{eq:avw}). If the CF is of the form 
$w_0(\theta) = A {\theta}^{1-\gamma}$ then its average can be calculated 
by numerical integration from (\ref{eq:avw}):
\begin{equation}
	\overline w_0(\theta) = 1.968\, A\, \theta^{1-\gamma} 
	= 1.968\, w_0(\theta). \label{eq:avw0}
\end{equation}
\vspace{1mm}

3.5. Variation of $w(\theta$) with the limiting magnitude\\[2mm]

By computing the CF from sub-catalogues with different limiting magnitudes
we can mimic the changes in depth. The limiting magnitude $m\,$was varied from 
$m=r_F$(max) = 15.5 mag up to $m$=18.5 mag in half magnitude steps. The 
result is shown in Figure 1.
The CFs on a broad, on a log--log plot more than two decades range 
are straight lines, i.e. $w(\theta, m) = A(m) \theta^{1 - \gamma}\,$. The
parameter in the exponent is  $\gamma = 1.69 \pm 0.02$. Using this value
of $\gamma$, we can measure the amplitudes of the straight part of the CF.
The parameters\\  log ($A(m)$)  deg$^{\gamma - 1})$ 
are given in the Table 1.\\[2mm]
\begin{center}
Table 1.\\[2mm]
\begin{tabular}{|c||c|c|c|c|c|c|c|}
\hline
$m$&15.5  &16.0 &16.5 &17.0 &17.5&18.0&18.5\\ \hline
log($A\,$ deg$^{\gamma-1})$ & -.64 & -0.79 & -0.95 & -1.12 & -1.28 & -1.465 &-1.63\\
\hline
\end{tabular}
\end{center}
\vspace{3mm}
In the 15.5 $\le m \le 18.5$ interval the amplitude of the CFs $A(m)$, 
measured from the horizontal part of the
function $w(\theta; m)/\theta^{1 - \gamma}\,$ can be parametrised  as
\begin{equation}
	A(m) = 0.0148\cdot 10^{0.33 \cdot (15.5 - m)}{\rm deg}^{\gamma - 1}. 
					                    \label{eq:parA}
\end{equation}
The amplitude changes a factor of ten in the three magnitude range, as
in MES or Postman et al. 1998.
The whole 2d CF, can be parametrised
as 
\begin{equation}
	w_{p}(\theta; m) = A_p(m) \theta^{1-\gamma_p}
	\left[1 +  \left({\theta_1 \over \theta}\right)^{\alpha}\right]^{-1}
  \left\{\left[1 + \left({\theta_2 \over \theta}\right)^{\nu}\right]^{-1} -
  \left[1 + \left({\theta_3 \over \theta}\right)^{\nu}\right]^{-1}\right\}. 
                                                    \label{eq:wpar}
\end{equation}
We use $\alpha = 9\,$ and $ \nu = 2.5$. The first bracket on the right hand 
side  describes the break--down of the scaling at small 
angles, while the break away from the power law  at large angular distances
contained in the second. In this formula we take into 
account that for small limiting magnitudes the 2d CFs are somewhat steeper,
and $\gamma_p\,$ varies slightly with the limiting magnitude, consequently 
$A_{p}(m)\,$ does not agree exactly with $A(m)\,$ given above, where we have
used the value $\gamma = 1.69$. The parameters 
in ({\ref{eq:wpar}) are given in  Table 2.\\
\begin{center}
Table 2.\\[2mm]
\begin{tabular}{|c||c|c|c|c|c|}
\hline
$m$ & $A_p$ deg$^{\gamma_p-1}$&$\gamma_p $ &$\theta_1\,$ deg 
&$\theta_2\,$ deg &$\theta_3\,{\rm deg}$\\
\hline \hline
15.5 &0.2352 &1.73 &8.8 10$^{-3}$& 6.5 & 40 \\ \hline
16.0 &0.1740 &1.71 &7.1 10$^{-3}$& 8.5 & 40 \\ \hline
16.5 &0.1350 &1.69 &5.5 10$^{-3}$& 8.0 & 40 \\ \hline
17.0 &0.0860 &1.68 &4.5 10$^{-3}$& 7.5 & 40 \\ \hline
17.5 &0.0550 &1.68 &4.1 10$^{-3}$& 7.0 & 40 \\ \hline
18.0 &0.0350 &1.68 &3.7 10$^{-3}$& 6.3 & 40 \\ \hline
18.5 &0.0245 &1.68 &3.4 10$^{-3}$& 6.0 & 40 \\ \hline
\end{tabular}
\end{center}
The maxima of the CFs are shifted toward smaller angles as the limiting 
magnitude grows. This corresponds to the change of the average apparent
galaxy diameter with $m$. When the apparent picture of the galaxies overlap,
the software identifies them in the ``unknown objects''. One can  parametrise
the maxima of the CFs as function of $m\,$ as 
$\theta_{\rm max} = 2.83 \cdot  10^{-{m \over 6.54}}$.

The slope of the straight-line part of the log($w$)--log($\theta$) function 
is between -1.73 and -1.68 in agreement with earlier results.  The functions 
with lower limiting magnitudes are somewhat steeper. This could mean that the 
intrinsically brighter galaxies are more correlated (Hamilton 1988, 
Davis et al. 1988, Benoist  et al. 1996).

Between 6$^o\,$ and 10$^o\,$ the CFs go through zero  (Figure 12 - 14).
The amplitude of the negative part decreases with increasing
magnitude limit.\\[2mm]
\newpage
3.6. Consistency of the two methods\\[2mm]

The CF measured by count in annuli method  and the average of the CF measured
by counts-in-cells for galaxies brighter than $r_F$ = 18 mag are shown in 
Figure 2. First of all we check whether the two methods are consistent by
computing the average of the CF determined by count in annuli method. 

The broken line in Figure 2. shows the the parametrised CF $w_p(\theta)$.
We compute the average of this function on squares with edge length
$\theta$. The direct Monte Carlo integration of $w\,$ would be dangerous
because of the peek at $\theta = 0$.  The average of $w_0(\theta)$ is known 
and the average of the difference $w_p(\theta) - w_0(\theta)\,$  is not 
singular any more and can be easily calculated from (\ref{eq:wpar}). The 
resulting averaged CF is shown by the 
dashed line in Fig. 2. The agreement of the two methods is excellent.\\[2mm]

3.7 Comparison with $w(\theta$,m) from the APM survey\\[2mm]

Maddox, Efstathiou and Sutherland 1996 have compared in detail the CFs, 
calculated from the APM survey with $w(\theta, m$) from other surveys. An 
important result of their analysis that large area surveys are required to 
determine $w$ reliably on angular scales of a few degrees. They found that the
deviation of $w$ from a power law seen in the small area catalogues 
are caused by the integral constraint in the estimator of $w$. 

The APM CF agrees well with that calculated from the Lick-survey (Groth and
Peebles 1977) for $\theta \le 3^o$, but the APM CF breaks less sharply from the
power law on scales larger than $\sim 3^o$. 
By comparing with the CF calculated from the Edinburgh--Durham Southern
Galaxy Catalogue (Collins et al. 1989, 1993) MES found that up to 2$^o$
the two CFs agree well but for larger angular distances the EDSGC gives more 
correlation than the APM catalogue.  

In Figure 3 we compare our CFs with that of MES (their Figure 23). The 
magnitude limit (after the colour correction $b_J - r_F$ = 1.5 mag) for the 
two upper curves on both panel are not exactly the same and this causes the 
small relative shift of the CFs. The agreement of the other pairs of curves  
are excellent. We confirm the results of MES concerning the $\theta \,$ 
and $m\,$ dependence of the 2d CFs. Note that the APM
CFs were calculated by using 80 - 120 Schmidt plates, the MRSS 152 plates; the 
plate sizes and their overlaps in the two catalogues are not the same, 
furthermore the N(b$_J$) and N($r_F$) functions are different,
the photometry and the star--galaxy separation software are independent etc.
In spite of all these differences the CFs measured from the two  catalogues 
agree in a range of several magnitudes. 
After the colour correction our CFs, determined from sub-catalogues with 
different limiting magnitude $m$, agrees excellently with 
the CFs  measured from the APM catalogue. For lower magnitude limits 
($m \le 17.0\,$mag) the MRSS functions show less correlation in the region 
$\theta > 4^o$. MES give the negative part of the 2d CF only for one magnitude
limit. Their CF shows much less anti-correlation as ours (Figure 4).\\[2mm]
\newpage
3.8. Scaling of the correlation functions\\[2mm]

Groth and Peebles 1977 have shown that the CFs of different $m_{lim}\,$
can be transformed into one curve by shifting them with appropriate amounts
$\Delta_x(m)\,$ and $\Delta_y(m)\,$ in the log($\theta)\,$ and log($w$)
directions.  
As it is shown in Figure 5 one can approximate the measured CFs on log--log 
scale with straight lines with slopes $\delta_1$ = -0.68 and 
$\delta_2$ = -3.0.
We could scale the section point into one points, in accord with
to the Groth -- Peebles 1978 prescription. When scaling the $m = 18$ mag
CF to  $m=17$ mag we must use $\Delta_x = 0.0055, \,\Delta_y = 0.332\,$ and 
for $m = 16$ mag CF to  $m=17$ mag the parameters $\Delta_x =-0.0038, \, 
\Delta_y =-0.231$. For the APM catalog MES obtained  a reasonable overlap 
of the transformed CFs. Using the EDSGC catalogue CNL92 demonstrated  
similar scaling.  
However, it will be shown by assuming the scaling of the CFs that the 
shifts in the log$(\theta$) and log($w$) directions are not independent from
each other. We discuss further the scaling properties in the section 4.2.

An other way to demonstrate the  scaling is to  divide the measured CFs by 
$A_p(m) \theta^{1-\gamma_p(m)}$. The resulting  function $ f(\theta$), is  
independent of the limiting magnitude $m$ in a large angular range, at least 
up to 6$^o$ (Figure 6).
The semilogarithmic presentation of the $f(\theta)$ function emphasises
the dip at about 0.4$^o$. A similar dip is seen in Figure 23 of MES96
and Figure 2 of CNL92. The reason for this dip seems to be obscure.

The CFs measured from MRSS and from the APM catalogue deviate  from the
power law at about $2^o$, independent of the limiting magnitude. The CFs 
determined from EDSGC break away from scaling at about 19$^o$, also 
independent of $m$ (CNL92 Figure 2).

The fact that the function $f(\theta)\,$ does not depend on $m$ in a bright
range of $\theta$ shows already that the 3d CF, calculated from the 2d CFs, 
must depend on the 
limiting magnitude $m$, i.e. on the depth of the catalogue.\\[2mm]

4. Relation between two dimensional and three dimensional correlation 
functions\\[2mm]

From a large surface deep 3d catalogue one could determine the luminosity
function of the galaxies, together with their luminosity evolution, as well
as the 3d correlation function and its evolution with the redshift. The 2d 
catalogues, like MRSS,
contain at least one order of magnitude more galaxies but the problems of the
selection function is not solved. In some sense the informations got from
the two type of catalogues complement each other. 

Here we determine from the measured 2d CF the $r_0\,$ and $\epsilon\,$ 
parameters of the  3d CF. In the literature there are several way to simplify
the problems. One possibility is to  take a $dN/dz\,$ function, measured from
a smaller catalogue, fix the parameter $\epsilon\,$ and calculate $r_0\,$ as 
a function of the magnitude limit. The second way is to fix $r_0\,$ and 
$\epsilon\,$ and choose appropriate parameters in the selection function.
A further method is to use a reliable $dN/dz\,$ function and to choose 
$\epsilon\,$ so that $r_0\,$ doesn't depend on the limiting magnitude (depth)
of the sub-catalogues.\\[2mm]
\newpage
4.1 The Limber-equation\\[2mm]

The  Limber-equation connects the 3d CF with the 2d CFs:
\begin{equation}
	 w(\Theta;m) = {1 \over \Theta} \int_0^{\infty} r 
	K(r/\Theta;m) \xi (r) dr                  \label{eq:wxi}
\end{equation} 
where $\Theta = 2\, {\rm sin}(\theta/2)$, and the kernel is
\begin{equation}
	K(t;m)= {2 \over (N_* \omega)^2} \int_0^t {F(x) (dN/dx)^2 \over 
	\sqrt{t^2 - x^2}} { dx \over (1 + z)^{3+\epsilon}}. \label{eq:kern}
\end{equation} 
Here $\omega$ is the solid angle covered by the survey, $N_*(m)$ is the number
of galaxies per steradian, $F(x) - 1\,$ is a relativistic correction (Peebles 
1980,\S 56), which is exactly zero for $\Omega_0$ = 1, and small for any 
$\Omega_0$ for the  depth covered in MRSS.
For the simplicity we use the Einstein-- de Sitter model ($\Omega_0 = 1, 
\Lambda$ =0). In this case the metric distance depends on the redshift as
\begin{equation}
	x = {2 c \over H_0} (1 - \sqrt{1 + z}).    \label{eq:xz}
\end{equation}
We use $H_0 =100 h$ km/s Mpc.

If the 3d CF is of the form  
$\xi(r) = \left({r_o \over r}\right)^{\gamma} $ then
from  (\ref{eq:wxi}) and  (\ref{eq:kern}) follows that $w(\theta ; m) =
A(m) \theta^{1 - \gamma}.$ The amplitude  $A(m)\,$ can be calculated from 
$\xi(r)$ and the normalised galaxy number distribution $p(x;m)$:
\begin{equation}
	p(x;m) = q(z;m) {dz \over dx} = {1 \over N_*(m) \omega} 
	{dN(z;m) \over dz} {dz \over dx};        \label{eq:p}
\end{equation} 
with
\begin{equation}
   \int_0^\infty p(x ; m) dx = \int_0^\infty q(z ; m) dz= 1. \label{eq:no}
\end{equation} 
The amplitude $A(m)\,$ dependent only on the form of the 
${dN \over dz}\,$ function and independent of its amplitude. If we know
$p(x;m)$, there is a unique connection between the correlation length
$r_0\,$ and the 2d amplitudes $A(m)$.\\[2mm]

4.2. Scaling and the Limber--equation\\[2mm]

Let us now assume that the 2d CFs scale with the limiting magnitude $m$, ie.
\begin{equation}
	w(\theta;m^{\prime}) = C w(\lambda \theta ; m).     \label{eq:wsc}
\end{equation} 
In this case from (\ref{eq:kern}) follows that $K(t;m)\,$ scales as 
\begin{equation}
	K(t ; m^{\prime}) = {C \over \lambda} K({t \over \lambda} ; m) 
                                                            \label{eq:Ksc}
\end{equation} 
and by (\ref{eq:p}) we have
\begin{equation}
	p(x ; m^{\prime}) = \sqrt{\lambda \over C} p(x/ \lambda ; m)
\left[{1 + z(x) \over 1 + z(x/\lambda)}\right]^{(3+\epsilon)/2}. \label{eq:psc}
\end{equation}  
From the normalisation condition of $p(x ; m^{\prime})\,$ follows that
\begin{equation}
	\lambda^{3 \over 2} C^{-{1 \over 2}} I(\lambda,m,\epsilon) = 1, 
						          \label{eq:Csig}
\end{equation} 
where
\begin{equation}
	I(\lambda, m, \epsilon) = \int_0^{\infty} \left[{1 + z(x \lambda) \over
	1 + z(x)}\right]^{(3 + \epsilon)/2} p(x;m) dx .    \label{eq:I}
\end{equation} 
That means that the shift of the $w(\theta;m)\,$ functions in directions
of the log($\theta)$ axis and log($w$) axis are not independent but
\begin{equation}
	\Delta {\rm log}(w) \equiv {\rm log}(C) = 3 \Delta {\rm log}
	(\theta) +2 {\rm log} (I(\lambda,m,\epsilon)).     \label{eq:shift}
\end{equation} 
For a given realistic distribution function $p(x; m$) or $q(z; m$) the integral
$I(\lambda, m, \epsilon$) can be calculated. It turns out that it gives
only a small contribution to $C$, corresponding to the fact that in the depth 
of MRSS the curvature effects are neglected and the redshift effects are small 
(see Groth \& Peebles 1977).  

Consequently,  if we scale together on the log--log plot the less steep part
of the CFs taking into account (\ref{eq:shift}), their steeper parts will
not fall together. That means that the correlation amplitudes and the position
of the break in $w$, as measured from the MRSS are independent, ie
log$[w(\theta;m)]$ can not be transformed into log $[w(\theta;m^{\prime})]$ 
by a shift of $\Delta {\rm log}(\theta)$ and $\Delta {\rm log}(w)$ given by 
(\ref{eq:shift}).
We conclude that if there exists a 3d CF, independent of the depth of the 
catalogue, the 2d correlation functions do not scale with the 
limiting magnitude $m$. The slope of the 2d CFs and the position of the
break down of the scaling are independent from each other.\\[2mm]

4.3 The selection function\\[2mm]

Maddox et al. 1990 proposed a selection function calculated from a
redshift dependent luminosity function. Later,  MES96  used
a selection function deduced from the measured redshifts of a part
of the galaxies in the 2d catalogue in the form of
\begin{equation}
	{dN \over dz} = N_* \omega  q(z;\alpha,\beta,z_0)  \label{eq:dNdz}
\end{equation} 
where $q(z;\alpha,\beta,z_0) $ was parametrised as
\begin{equation}
	q(z;\alpha,\beta,z_0) = {\beta \over \Gamma \left({\alpha +1
	\over \beta} \right) z_0} \left({z \over z_0}\right)^{\alpha}
	e^{-({z \over z_0})^{\beta}}. \label{eq:f}
\end{equation}
with $\alpha$ = 2 and $\beta$ = 3/2. The maximum
of this distribution is z$_{max} = (\alpha / \beta)^{1/\beta} z_0$,
the mean value of the $n$-th moment of the redshift $\overline {z^n(m)}$ is 
\begin{equation}
	\overline {z^n} = z_0 {\Gamma((\alpha +1+n)/\beta) \over 
	\Gamma((\alpha +1)/\beta)} \label{eq:avz}
\end{equation}
and the modal value can be calculated from the following equation
\begin{equation}
	\Gamma({\alpha +1\over \beta},z_{mod}) = {1 \over 2}
	 \Gamma({\alpha +1\over \beta}).     \label{eq:modz}
\end{equation}
For $\alpha$ = 2 and $\beta$ = 1.5 one has $z_{max} = 1.2114 z_0$,
$\overline z$ = 1.5048 z$_0$ and z$_{mod} = 1.412 z_0.$ From these $dN/dz\,$
function one can determine the luminosity function. It  depends mildly on $z$. 
This parametrisation is simple and general enough to describe the selection
function. 

We use two set of parameters $\alpha, \beta\, {\rm and}\, z_0$.
First we take, following MES96  $\alpha = 2\, \beta =1.5\,$ and determine 
$z_0\,$ from the condition $r_0 = 5 h^{-1}$ Mpc (and $\epsilon$ = -1.3). The 
second set of parameters was calculated by using the luminosity function,
determined from the Las Campanas survey (Lin et al. 1996). This must  better 
approximate the real selection function, because the $R\,$ band 
used in the Las Campanas survey is not far from the $r_F\,$ band of the MRSS.
The parameters are given in Table 3.\\[2mm] 
\centerline {Table 3. Parameters of the selection function}
\vspace{2mm}
\begin{center}
\begin{tabular}{|c||c|c|c||c||c|c|c|c|}
\hline
	&\multicolumn{3}{c||} {Parameter set 1}&\multicolumn{1}{c||}{APM}
                   &\multicolumn{4}{c|}  { Parameter set 2}\\
     	&\multicolumn{3}{c||} {$ r_0 = 5 h^{-1}$ Mpc}
                   &\multicolumn{1}{c||} {$b_j<m$+1.5}
                   &\multicolumn{4}{c|}  { From Las Campanas LF}\\  \hline
$m$ & $\alpha$ &$\beta$ &$z_0\,$&$z_0$& $\alpha$ &$\beta$ &$z_0\,$&$\overline{z}$\\
\hline \hline
15.5 &2 &1.5&0.0266&0.0326&2    &1.50  &0.0284&0.0436\\ \hline
16.0 &2 &1.5&0.0327&0.0366&2    &1.50  &0.0350&0.0527\\ \hline
16.5 &2 &1.5&0.0414&0.0439&2    &1.52  &0.0440&0.0651\\ \hline
17.0 &2 &1.5&0.0516&0.0534&2    &1.57  &0.0568&0.0804\\ \hline
17.5 &2 &1.5&0.0645&0.0646&1.95 &1.62  &0.0730&0.0965\\ \hline
18.0 &2 &1.5&0.0834&0.0774&1.90 &1.66  &0.0921&0.1205\\ \hline
18.5 &2 &1.5&0.1050&0.0915&1.90 &1.695 &0.1140&0.1461\\ \hline
\end{tabular}
\end{center}
\vspace{4mm}
In the second parameter set the parameter $\beta\,$ grows with the depth of 
the catalogue. This fits in the general picture. For deep 
($\overline{z} > 0.3$) 
catalogues several author (Lilly et al. 1995, Brainerd \& Smail 1998) use the 
same function form with $\beta$ =  2, or modifications of it( Postman et al. 
1998). We note already here that our main results do not depend essentially 
on the  exact form of the selection function.\\ 

4.4. Solution of the Limber-equation for CF of power law form\\[2mm]

When the 3d CF follows a power law and the normalised selection function is 
known, ie. the parameters 
$\alpha, \, \beta$ and $z_0 \,$ are given, there is a simple relation between
$A(m)\,$ and $r_0\,$ (Totsuji et Kihara 1969). For small $r$ the 3d CF is 
$\xi(t \Theta) = \left({r_0 \over t \Theta}\right)^{\gamma}$, and we get
\begin{equation}
	A(m) = \Theta^{\gamma - 1} w(\Theta ; m) = r_0^{\gamma} 
	\int_0^\infty t^{1 - \gamma} K(t;m) dt.         \label{eq:Aam}
\end{equation}
By using (\ref{eq:xz}) and developing ${dz \over dx} (1 + z)^{-(3+\epsilon)}$
by $z$ up to the second order, we can compute the integral (\ref{eq:Aam}):
\begin{equation}
	A(m) = C  \left({H_0 r_0 \over c z_0 }\right)^{\gamma}
	\left[ \Gamma\left(\tau\right) + 
	C_1  \Gamma\left({\tau - 1/ \beta}\right) z_0 + 
	C_2  \Gamma\left({\tau - 2/ \beta}\right) z_0^2\right] \label{eq:segg}
\end{equation}
where
\begin{equation}
	\tau = {2 +2 \alpha -\gamma \over \beta}, \qquad\,\,\,  
	\qquad \,\,\,\, C = 2^{1 - \tau} {\beta \, \sqrt {\pi} \,
	\Gamma ({\gamma - 1 \over 2}) \over {\Gamma}^2 
	({ \alpha +1 \over \beta}) \,\Gamma({\gamma \over 2})} \label{eq:cons}
\end{equation}
and
\begin{equation}
  C_1 = 2^{-{1 \over \beta}} {-9+3 \gamma - 4 \epsilon \over 4};
	\,\,\,\, C_2 = 2^{-{2 \over \beta}} {9 \gamma^2 -65 \gamma +
    16 \epsilon^2 + 88\epsilon - 24 \gamma \epsilon + 116 \over 32}.
                                                                \label{eq:C}
\end{equation}
With the selection function, calculated from the Las Campanas luminosity  
function (Lin et al. 1996) we can determine from the measured $A(m)\,$  using  
(\ref{eq:segg}) the parameter $r_0$:
\begin{equation}
	r_0 = {c \over H_0} z_0 A(m)^{1 \over \gamma} 
	C^{-{1 \over \gamma}} \left[\Gamma(\tau) +
	C_1 \Gamma(\tau - 1/ \beta) z_0 + 
	C_2 \Gamma(\tau - 2/ \beta) z_0^2 \right]
	^{-{1 \over \gamma}}.                                   \label{eq:r0Am}
\end{equation}

If we fix the correlation length $r_0\,$ and $\epsilon$,
the equation (\ref{eq:segg}) can be solved for $z_0$ by iterations. 
First we neglect the terms containing $z_0$  in the square bracket. Then
\begin{equation}
	z_0^{(0)} = A(m)^{-{1 \over \gamma}} {H_0 \, r_0 \over  c} 
	\left[ C \Gamma \left({\tau}\right) \right]^{1 \over \gamma}.
                                                      \label{eq:z00A}
\end{equation}
substituting that into (\ref{eq:segg}) one gets
\begin{equation}
	z_0^{(1)}(m) = z_0^{(0)}\left[ 1 + C_1{\Gamma(\tau - 1/\beta) \over 
	\Gamma({\tau})} z_0^{(0)} + C_2 {\Gamma(\tau - 2/\beta)  \over 
	\Gamma({\tau})} {z_0^{(0)}}^2\right]^{1 \over \gamma}.   \label{eq:z0A}
\end{equation}
For $r_0 = 5 h^{-1}$ Mpc and $\epsilon$ = -1.3
$z_0(m)$ is shown in Table 3. and can be parametrised as 
$z_0(m) = 0.024 + 0.0117 (m - 15.5)^{1.63}.$   
Of course, it is not granted that these $z_0\,$ values correspond to the real 
$dN/dz\,$ distribution. This parametrisation agrees well with that of  MES for
$m\sim$ 17 mag, but our $z_0(m)\,$ changes somewhat faster with $m$.  The same
is true for out second parametrisation.

Next, with the second parameter set we compute $r_0(m)$, for $\epsilon = -1.3$.
In Figure 7. we show our $r_0(m)\,$ function with  and compare it with the 
same function, calculated from the data given in MES (their formula 38b and 
Figure 26a) for the same $\epsilon$. The correlation length $r_0\,$ depends 
in both calculation on the limiting magnitude.  The redshift dependence of 
the CF is beside of the factor $(1 + z)^{-3 - \epsilon}$ also in $r_0$.

It is possible to concentrate the redshift dependence of the correlation 
function in the $(1+z)^{-{3+\epsilon \over \gamma}}\,$ factor. By minimising 
the change of $r_0(m)\,$ with the limiting magnitude $m$, we get 
\begin{equation}
	r_{00} = 5.82 \pm 0.05 \vspace{10mm} \quad {\rm and}\quad 
	 \epsilon = 2.60.                            \label{eq:r0e}
\end {equation}
This result shows that the correlation length $r_0({\overline z})$ decreases 
rapidly with the average redshift of the sub-catalogue. It means that the 
stable clustering is not  yet reached, on small scales there is an inflow in 
physical coordinates. 

We estimate the average inflow/outflow velocities from the continuity equation
as follows:
\begin{equation}
   	v = H(z)\, r\, {1 - {\epsilon \over 3 - \gamma} \left({r_0(z) \over r}
	\right)^{\gamma} \over 1 + \left({r_0(z) \over r}\right)^{\gamma}}.
\end {equation}
If $r \gg r_0\,$ v corresponds to the Hubble-flow and if $ r \ll r_0,$ then
there is an inflow with $ v \approx -{\epsilon \over 3 -\gamma} H r$. This  
makes possible e.g. to estimate the rate of galaxy merging.\\[2mm]
 
4.5 The kernel of the integral (\ref{eq:wxi})\\[2mm]

In the following sections we determine at which $r\,$ breaks the 3d CF down
from the power law, by using the 2d CFs, measured from different sub-catalogues.
To relate the 2d and 3d correlation functions of general form we must now 
determine the (\ref{eq:kern}) kernel. The integral is easily calculated.
The $Q(t,m) = t^{1 - \gamma} K(t;m)$ functions are shown in Fig.8 for 
$15.5 \le m \le 18.5$ mag. We included in $Q(t,m)\,$ the 
multiplier $t^{1 - \gamma}$, so the remaining factor under the integral is 
constant up to the point, where the 3d CF breaks down from scaling . With 
growing magnitude limit $m$, the maxima of the functions $Q(t,m)\,$
shifted towards higher $t = r \theta$, they become flatter and in 
the integral (\ref{eq:wxi}) the large $t$ part of the 3d CF will be more 
important.

The  $Q(t,m)\,$ functions have a very simple scaling. As it is to see from the
non-relativistic, curvature free approximation of (\ref{eq:kern})
\begin{equation}
	Q(t,m)= b\, Q(t/a;m^{\prime}).         \label{eq:scal}
\end {equation}
The coefficients for  scaling on $Q(t;17{\rm mag})$ are given in 
Table 4.\\[2mm]
\centerline {Table 4. Scaling factors to 17 mag}
\vspace{2mm}
\begin{center}
\begin{tabular}{|c||c|c|c|c|}
\hline
$m$ & $a$ &$b$&log$ a$ & log $a^{\gamma} b$\\ \hline \hline
15.5 &1.815 &0.187& 0.2589 &-0.2883\\ \hline
16.0 &1.530 &0.302& 0.1847 &-0.2079\\ \hline
16.5 &1.245 &0.545& 0.0952 &-0.1028\\ \hline
17.0 &1.00  &1.00 & 0.     &  0.   \\ \hline
17.5 &0.844 &1.655&-0.0731 &-0.0952\\ \hline
18.0 &0.690 &3.141&-0.1612 & 0.2245\\ \hline
18.5 &0.578 &5.467&-0.2381 & 0.3354\\ \hline
\end{tabular}
\end{center}
\vspace{4mm}
Notice that $b \approx a^{-3}$. The scaled functions are shown in Figure 9.
By taking a 3d CF, independent of the sample depth, one can deduce from 
(\ref{eq:scal}) a scaling law for the 2d CFs:
\begin{equation}
	w(\theta,m)= {1 \over b a^{\gamma}} w(a \theta; 17).   \label{eq:scalw}
\end {equation}
In Figure 10. we show the 2d CFs measured from sub-catalogues $m$ = 15.5 mag 
and
$m$ = 18.5 mag scaled to $w(\theta$; 17 mag) by using (\ref{eq:scal}).
One sees again that if the 3d CF independent  from the depth of the catalogue,
the large $\theta\,$ part of the 2d CFs do not scale with the limiting 
magnitude. \\[2mm]

4.6 Modelling of the 3d correlation function\\[2mm]

Baugh 1996 solved the Limber-equation for $ \xi(r)\,$ by inverting 
(\ref{eq:wxi}) through an iterative method. He used the full 
$17 \le b_J \le 20\,$ slice of the APM catalog. The resulting 3d CF 
(for $\epsilon = -3$) well fitted by the $(4.5 {\rm Mpc}/r)^{1.7}$
power law for $r \le 4 h^{-1}\,$ Mpc, between $4 h^{-1} \le r \le 25$ Mpc
$\xi(r)\,$ has a shoulder and rising over the quoted power law. 
For $r > 25 h^{-1}$ Mpc the calculated 3d CF becomes negative and on the 
scales $r > 40 h^{-1}$ Mpc consistent with zero. Baugh 1996 calculates also 
the 3d CF for other magnitude slices and finds small discrepancies between 
these functions  recovered from various magnitude slice of the APM catalogue.

Here we use another method. We parametrise the 3d CF in different forms and 
compute the 2d CFs for several magnitude limits $m\,$ from the 
Limber--equation (\ref{eq:wxi}). The best fitting form and parameters
determine the 3d correlation function. It is clear that the form of the 3d CF 
must be the same type as that of the 2d CFs. To describe the physics mirrored
by the correlation function, several distance like parameter are used: $r_0\,$ 
gives its amplitude for small $r$. This parameter is simple connected with
the amplitude of the 2d correlation function (\ref{eq:segg}). It is also clear 
that the 3d CF must cross zero at some $r_2$, to permit the change of sign of 
the 2d CFs. It is generally assumed that there in no correlation between the 
galaxies at distances large enough ie for $r \rightarrow \infty$  the 
correlation function $\xi \rightarrow 0$. This asymptotic is characterised by 
a parameter $r_3$. Further, by considering some concrete form of the 3d CF
$\xi(r; r_1,r_3)$, one can not reconstruct with (\ref{eq:wxi}) the elbow of the
2d CF. We must introduce a further parameter in $\xi$ what describe the 
``shoulder'' seen by Baugh 96. We think, this shoulder corresponds to the 
extra correlation of the galaxies because of the cluster - cluster correlation.

If we use the 
integral constraint $\int r^2 \xi(r) dr = 0$, we get a relation between the 
parameters. We have tried numerous function form. Here we describe only one 
parametrisation, since our main result, described in the next subsection do 
not depend sharply on the details of the parametrisation. Our best 
function has the form
\begin{equation}
       \xi(r) = \left[\left({r_0 \over r} + {r_0 \over r_1}\right)^{\gamma}
	-\left({r_0 \over r_1} +{r_0 \over r_2}\right)^{\gamma}\right] 
	e^{-r/\lambda}.                      \label{eq:xipar}
\end{equation}  
As it was shown in the previous section, one can not fit all the 2d CFs with 
a magnitude limit independent 3d CF. \\[2mm]

4.7 Three dimensional correlation function determined from 
different 2d sub-catalogues\\[2mm]

The parameters $r_0\,$ and $\epsilon\,$ of the 3d CF are fixed by the
scaling part of the 2d CFs and are given in (\ref{eq:r0e}).
Now we determine the other parameters of the 3d CF $\xi(r)\,$in 
(\ref{eq:xipar}). For $m$= 17 mag we get $r_1$= 8.47 h$^{-1}$ Mpc,
$r_2$= 39.95 h$^{-1}$ Mpc and  $\lambda$= 41.70 h$^{-1}$ Mpc. The 3d CF 
determined by these parameters is not far from the correlation function
computed by Baugh 1996 from tha APM catalogue. However, if we use the same 
parameters for sub-catalogues with other magnitude limit, we find that 
for $m < 17$ mag the 2d CFs, calculated from (\ref{eq:wxi}) breaks down from
scaling later than measured, for $m > 17$ mag earlier. We find that the 
3d CF, calculated from different sub-catalogues, changes in a systematic manner.
To stress that, we parametrise $r_1,r_2\,$ and $\lambda\,$ as follows
\begin{eqnarray}
	r_1       =  7.50 + 0.65 (m - 15.5) h^{-1} {\rm Mpc} \nonumber\\ 
	r_2       =  25.1 + 9.92 (m - 15.5) h^{-1} {\rm Mpc} \nonumber\\
	\lambda\, =  31.2 + 7.98 (m - 15.5) h^{-1} {\rm Mpc} \nonumber\\
	\lambda_0 =  13.3 + 5.10 (m - 15.5) h^{-1} {\rm Mpc}
                                           \label{eq:param}
\end{eqnarray}  
The 3d CF determined from subcatalogues with different limiting magnitude 
$m\,$ are shown in Figure 11. 
The 3d CF determined by the parameters $r_1,r_2\,$ and $\lambda\,$ do not 
satisfy the integral condition. To fulfil it, we should use in  the last 
factor of (\ref{eq:xipar}) $\lambda_0$ instead of $\lambda$, but it would give
 a bad fit for the negative part of the 2d CFs (the tail of the  calculated 
2d CF would be much less negative). It could be explained by the 
oscillation of the large $r\,$ part of the 3d CF  as in Tucker et al. 1997.

The quality of the parametrisation and the choice of the parameters can be 
judged from the Figures 12 -- 14. We have tried other $dN/dz\,$ functions, 
other parametrisation for $\xi$, other $\epsilon\,$ value (with depth dependent
$r_0$) and we found always that the 3d CF, calculated from the measured 2d CFs 
depend on the catalogue depth.\\[2mm]

5. Summary and discussion\\[2mm]

We determined the two point correlation function from the Muenster Red Sky
Survey (MRSS, Ungruhe 1999). This survey uses the F--band and larger than the 
APM survey (MES). Maddox \& al.1990 and MES have compared their measured
correlation functions with those, determined from other catalogues. Here we
compare our 2d CFs with the results of MES.

The large dynamics range of the survey made it possible
to determine the correlation function from subcatalogues with different
magnitude limit $m$. The measured slope of the 2d CFs is somewhat larger for 
lower magnitude limit, but its average value 
$\delta =  \gamma - 1 = 0.69 \pm 0.02$, argees  with the earlier measurements.
The $\theta_{max}(m)$ value, where the CFs measured from subcatalogues with
different magnitude limits $m$ correspond to about 30$h^{-1}$ kpc projected 
distance, independent of $m$. When the projected distance of two galaxies is
smaller than that, the software can not separate the two objects. The 
amplitude of the correlation functions, after the color correction, agrees 
excellently with those measured from the APM catalogue. As it is shown in
Figure 3.  our CFs agree well with those, measured by MES also in the region
where the CFs do not follow the power law. Note that the break away from the 
power law begins at about the same angle, almost independent of the magnitude 
limit of the subcatalogue. After the zero crossing the MES correlation 
function is zero. We find that the amplitude of the negative part of the 
CFs is several times 10$^{-3}$, and decreasing with increasing magnitude
limit $m\,$ (Fig. 12 - 14). The non-vanishing anticorrelation is a natural
consequence of the definition of the correlation function.

We have calculated from the measured 2d CFs the three dimensional two-point 
correlation function by using the Limber-equation.
To determine the $dN/dz\,$ function we used informations from 3d
catalogues. We parametrizer it in the usual form and the parameters were
for each subcatalogue deduced  from the luminosity function of the Las 
Campanas survey (Lin et al. 1996). 
As it is shown in Fig. 9, the kernel of the integral in the 
$w(\theta) -- \xi(r)\,$ transformation (deprojection) scales with the
limiting magnitude of the subcatalogue.

From a sub-catalogue with limiting magnitude $m$ one can only determine a 
combination of $r_0(m)\,$ and $\epsilon$. There are several possibility
to separate these parameters. The usual way is to fix $\epsilon\,$
on the basis of physical considerations or simple ignore it. To compare our 
parameters with the $r_0(m)\,$ and $\epsilon$, determined from the APM 
catalogue, we used $\epsilon = -1.3$ and calculated $r_0(m)\,$ from the
subcatalogues. The agreement (Fig. 7) is excellent.
However, if one considers $\epsilon\,$ as evolution parameter it is more 
consequent to choose the correlation length $r_{00}\,$independent from the 
depth of the subcatalogue, ie. 
$r_0(\overline{z}) = r_{00} (1 + \overline{z})^{-{3 + \epsilon \over \gamma}}$
With this parametrisation we have got $r_{00} =5.82 \pm 0.05 h^{-1}{\rm Mpc
and} \epsilon = 2.60$. This result agrees with the general trend: deeper the 
catalogue lower the correlation length $r_0(\overline{z})$.  This 
$\epsilon\,$ value is unusually high. That means that there is yet an 
evolution in clustering.  
It is not easy to compare our $r_0, \epsilon\,$ with that of other authors, 
because the separation of the two parameters is not standardised.  From the 
Las Campanas survey Jing, Mo \& B\"orner 1998 found
$\gamma$=1.86 and $r_0$=5.06 h$^{-1}$ Mpc. With $\epsilon = \gamma -3\,$ and
$\overline {z}$ =0.1 that gives $r_0\approx$ 4.6 Mpc, in agreement with our
$5.82\cdot 1.1^{-5.6/1.68}$ = 4.2 Mpc. The limiting magnitude of the EPS 
catalogue (Guzzo \& al., 1999) is b$_J<$ 19.4 mag; they found  $\gamma$ = 
1.67 and $r_0=4.15\pm 0.2 h^{-1}$Mpc. We get for $r_F<18\,$mag $r_0$ = 3.99
$^{-1}$Mpc. For deep surveys it is questionable if the parametrisation 
(\ref{eq:r0z}) correct. Interestingly, extrapolating our parametrisation 
(\ref{eq:r0e}) to high $z\,$ values we are not far from the measured values.
E.g. $z=0.5\,$ we get $r_0 =$ 1.4 h$^{-1}$ Mpc (in physical coordinates, 
$q_0$=0.5). This value agrees well with the result of Le F\'evre \& al.1996, 
($r_0=$1.33h$^{-1}$ Mpc).  Our $r_{00}, \epsilon\,$ values  are compatible 
with the result of Postman \& al. 1998, who find that if $r_0\,$ is fixed to 
be 5.5$\pm$1.5 h$^{-1}$Mpc, then -0.4<$\epsilon$<+1.3. Connolly \& al. 1998
measure from a subcatalogue of HDF for fixed $r_0 = 5.4 h^{-1}$Mpc 
$\epsilon = 2.37 ^{+0.37}_{-0.64}$.

By comparing our 3d CF determined from the $r_F$<18.5 mag catalogue with that
of Baugh 1996 for 17 mag <$b_J$<20 mag, we see three minor differences. First, 
our correlation length $r_0\,$ (after taking into account the corrections for 
the different $\epsilon$'s) is about a factor of 1.3 larger. It can be 
explained by the difference of the $z_0\,$ parameters of the $dN/dz\,$
function. Second,  our correlation function crosses zero at 50 h$^{-1}$ Mpc, 
that of Baugh at 40 h$^{-1}$ Mpc. 
Third, the CF of Baugh 1996 disappears at about 80 h$^{-1}$ Mpc, ours seems to
be negative also for larger distances. This is a consequence of the difference
of the measured 2d CFs in this region. 

As it is shown in Fig. 11 and  (\ref{eq:param}) the position of the break-away
from the scaling of the 3d CF and the zero--crossing varies with the depth of 
the sub-catalogues.  This agrees with the result of Cappi et al. 1998. We
conclude that the MRSS is not deep enough ($z_{med} \approx$ 0.14) to 
determine up to which physical distance follows the 3d CF the power law.
The structures, found in large and deep (Sloan and 2dF) catalogues  are much  
larger then those seen in the MRSS. 

The author thanks all members of the Astronomical Institute of the University
Muenster, especially  Weltraud Seitter and Renko Ungruhe providing an 
excellent galaxy survey. I thank also George Efstathiou and Alex Szalay
for valuable discussion.
\newpage

References\\
Baugh C.M. 1996, MNRAS 280, 267\\
Baugh C.M. and Efstathiou G. 1993, MNRAS 265, 146\\
Brainerd T.G. and Smail I., 1998, ApJL, 494, 137\\
Carlberg R.G., Cowie L., Songalia A. and Hu E. 1997, ApJ, 484, 538\\
Conolly A.J., Szalay A.S. and Brunner R.J., 1998, ApJ, 499, L125\\
Davis M. and Peebles P.J.E., 1983, ApJ, 267, 465\\
Groth E.J., Peebles P.J.E. 1977, ApJ, 217, 385\\
Guzzo L., Iovino A., Chincarini G., Giovanelli R., and Haynes M. P. 1991\\
ApJ 382, L5\\
Guzzo L., Bartlett J., Cappi A. et al. 2000 AA 355 1
Jing Y.P., Mo H.J., B\"orner G., 1988, ApJ, 494, 1\\
Jones L.R., Fong R., Shanks T., Ellis R.S. and Peterson B.A. 1991 
MNRAS 265, 146\\
Joyce M., Montuori M. and Sylos Labini F. 1999 ApJ 514, L5\\
Joyce M., Anderson P.W., Montuori M., Pietronero L. and Sylos Labini F. 2000
Europhysics Lett. 50 (3) 416\\
Le F\'evre O.,Hudon D.,Lilly S., Crampton D., Hammer F. and Trees l., 1996,
ApJ, 461, 534\\
Limber D.N., 1953, ApJ, 119, 655\\
Maddox S.J., Efstatiou G. and Sutherland W.J., 1996,  MNRAS, 283, 1227\\
Maddox S.J., Efstatiou G., Sutherland W.J. and Loveday J. 1990, 
MNRAS, 242, 43p\\
Maddox S.J., Efstatiou G., Sutherland W.J. and Loveday J. 1990, 
MNRAS, 243, 692\\
Marzke R.O., Geller M.J., da Costa L.N. and Huchra J.P. 1995, AJ, 110, 477\\ 
Neuschaefer L.W., Im K., Ratnatunga K.U., Griffiths R.E., and Casertano S.
1997\\,ApJ 480, 59\\
Postman M., Lauer T.R., Szapudi I. and Oegerle W., 1998, ApJ, 506, 33\\
Schuecker P. 1990, PhD University of Muenster\\
Schuecker P. 1993, ApJ Suppl. 84, 39\\
Small T.A., Ma C.p., Sargent W.W. and Hamilton D. 1999, ApJ 524, 31\\
Spiekermann G., 1994 Rev. Mod. Astron. 7, 1176\\
Stevenson P.F.R., Shanks, T., Fong, R. \& MacGillivray, H.T., 1985,
Mon.Not. R. astr. Soc. 213 953 \\
Shechtman, S.A., Landy S.D., Oemler A. Jr., Tucker D.L., Lin H., Kirshner R.P.
and Schechter P. 1996, ApJ 470,172\\
Shepherd C.W., Carlberg R. G., Yee H. and Ellington E. 1997, ApJ, 479, 82\\
Sylos Labini F., Montuori M. and Pietronero L. 1998,  Physics Reports 293, 61\\
Szokoly G.P., Jain B., Budav\'ari T., Connolly A.J. and Szalay A.S., 2000\\
Tucker D.L., Oemler A. Jr., Kirshner R.P., Lin H., Shechtman S.A., Landy S.D.,
Schechter P.L., M\"uller V., Gottl\"ober S. and Einasto J. 1997, 
MNRAS, 28, L5\\
Woods D. and Fahlman G.G.1997 ApJ, 490, 11\\

\newpage
\begin{center}
\begin{figure}
\leavevmode \epsfxsize=5in \epsfbox{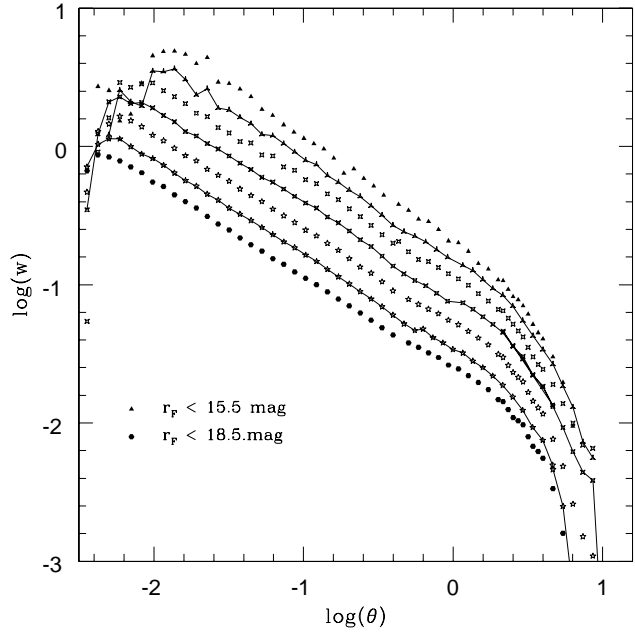}
\caption{\small Measured CFs for seven limiting magnitude from $r_F < $15.5
 to $r_F < 18.5$ derived from 152 plates}
\end{figure}
\end{center}

\newpage
\begin{center}
\begin{figure}
\leavevmode \epsfxsize=5in \epsfbox{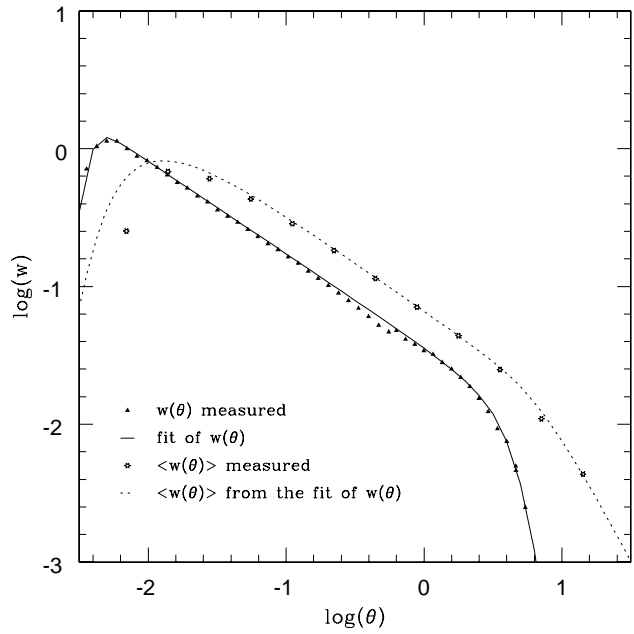}
\caption{\small Correlation function and the average 
of the correlation function}
\end{figure}
\end{center}

\newpage
\begin{center}
\begin{figure}
\leavevmode \epsfxsize=5in \epsfbox{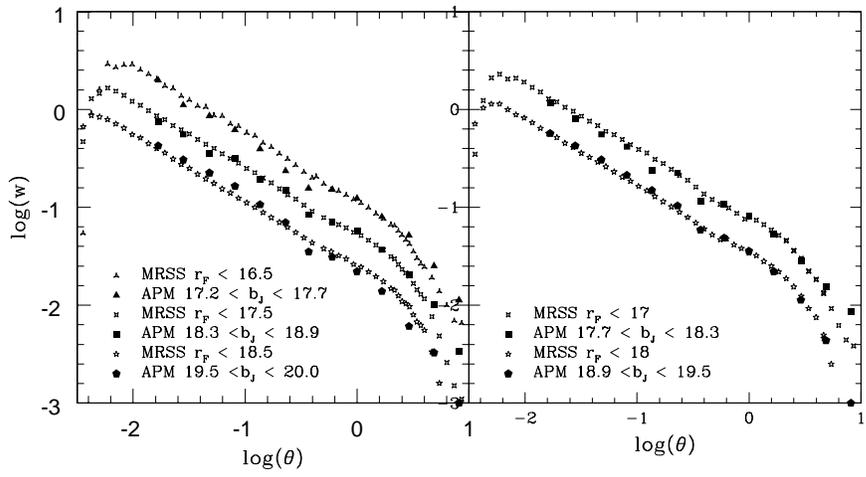}
\caption{\small Correlation functions as measured from MRSS
(small symboles) and from APM (large symboles) in MES Fig 23.}
\end{figure}
\end{center}

\newpage
\begin{center}
\begin{figure}
\leavevmode \epsfxsize=5in \epsfbox{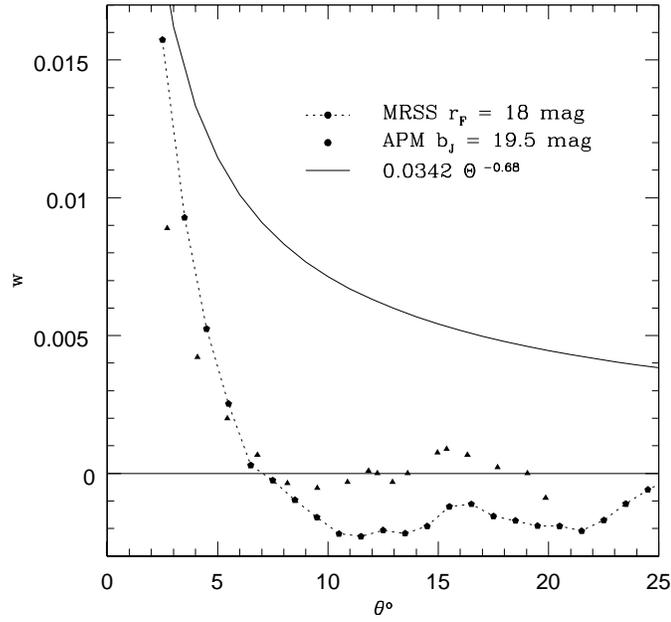}
\caption {\small Comparison the APM and MRSS correlation function 
on linear scale}
\end{figure}
\end{center}

\newpage
\begin{center}
\begin{figure}
\leavevmode \epsfxsize=5in \epsfbox{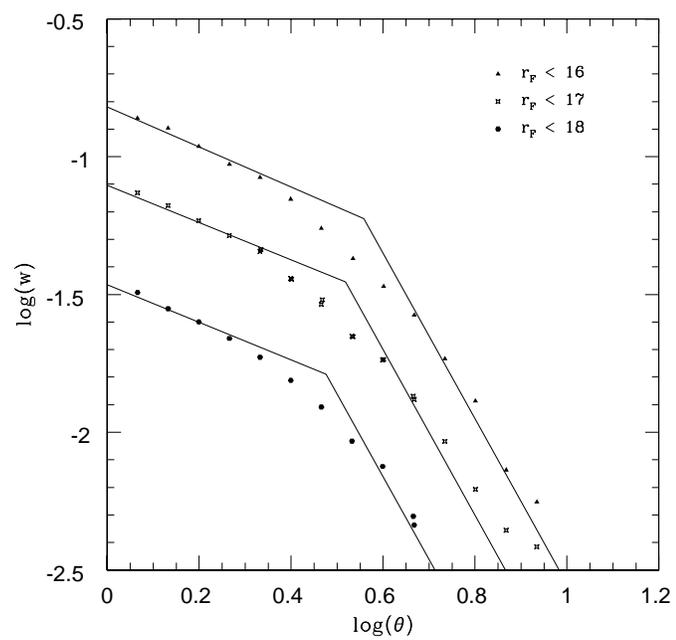}
\caption {\small Approximation of the CFs with double power law}
\end{figure}
\end{center}

\newpage
\begin{center}
\begin{figure}
\leavevmode \epsfxsize=5in \epsfbox{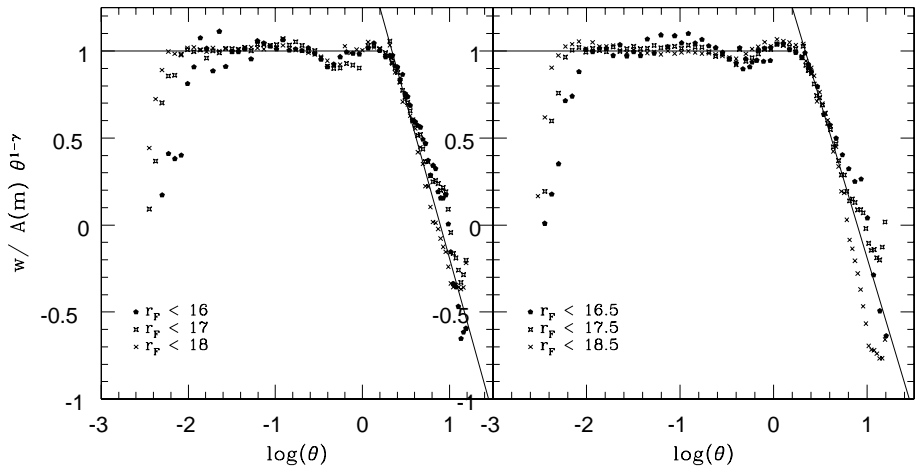}
\caption{ \small The function 
$f(\theta) = {w(\theta;m) \over A(m) \theta^{1 - \gamma}}$}.
\end{figure}
\end{center}

\newpage
\begin{center}
\begin{figure}
\leavevmode \epsfxsize=5in \epsfbox{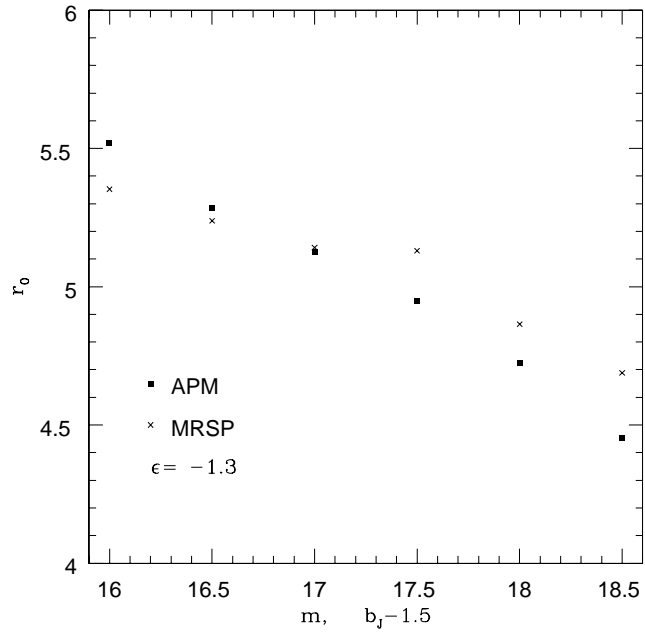}
\caption{ \small Correlation length $r_0\,$ in function of the limiting 
magnitude calculated from the APM catalogue and from MRSS for $\epsilon$=-1.3 }
\end{figure}
\end{center}

\newpage
\begin{center}
\begin{figure}
\leavevmode \epsfxsize=5in \epsfbox{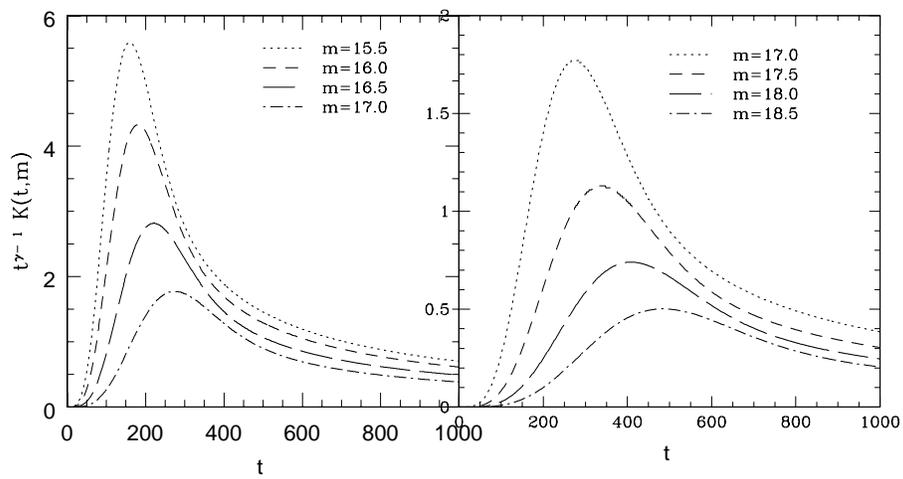}
\caption{ \small The kernel of the integral (\ref{eq:kern})
for different magnitude limits.}
\end{figure}
\end{center}

\newpage
\begin{center}
\begin{figure}
\leavevmode \epsfxsize=5in \epsfbox{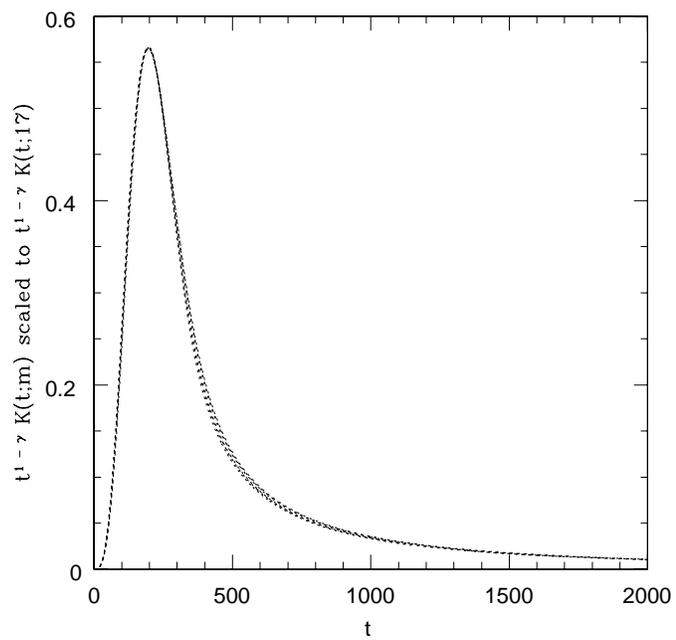}
\caption{ \small The function $t^{1 - \gamma} K(t;17)$ and 
the other six $t^{1 - \gamma} K(t;m \ne 17)$ functions, scaled on it
with (\ref{eq:scal}).}
\end{figure}
\end{center}

\newpage
\begin{center}
\begin{figure}
\leavevmode \epsfxsize=5in \epsfbox{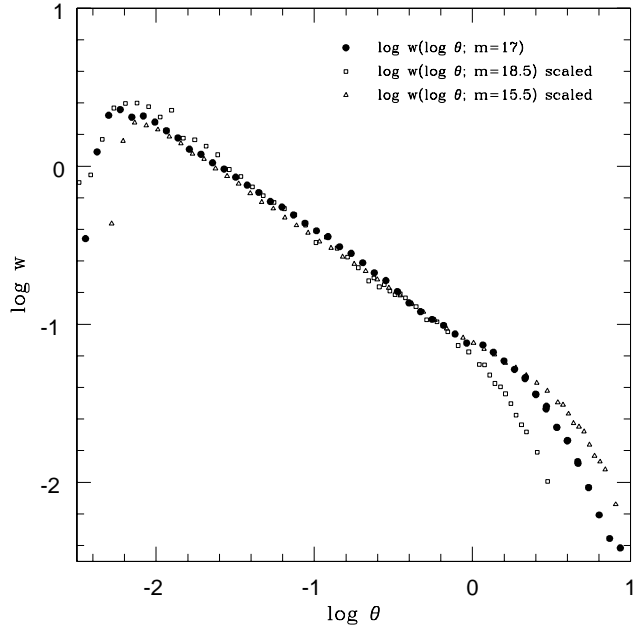}
\caption{ \small The 2CF, measured from a subsample $r_F \le 17$ mag
and the 2CFs log $w(\theta;15.5)$ \&  log $w(\theta;18.5)$ scaled on it 
with (\ref{eq:scalw}).}
\end{figure}
\end{center}

\newpage
\begin{center}
\begin{figure}
\leavevmode \epsfxsize=5.5in \epsfbox{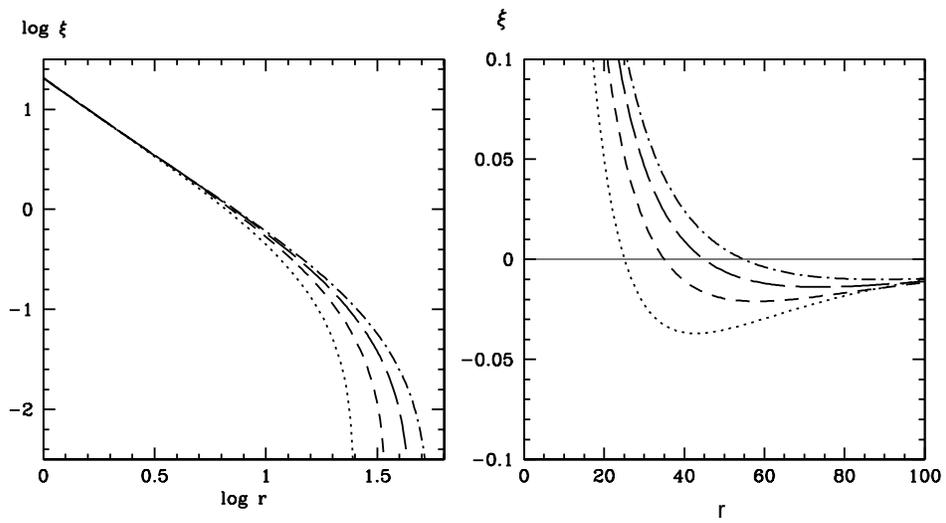}
\caption{ \small 3d correlation function determined from the subcatalogues
with limiting magnitudes 15.5, 16.5, 17.5 and 18.5 mag.}
\end{figure}
\end{center}

\newpage
\begin{center}
\begin{figure}
\leavevmode \epsfxsize=5in \epsfbox{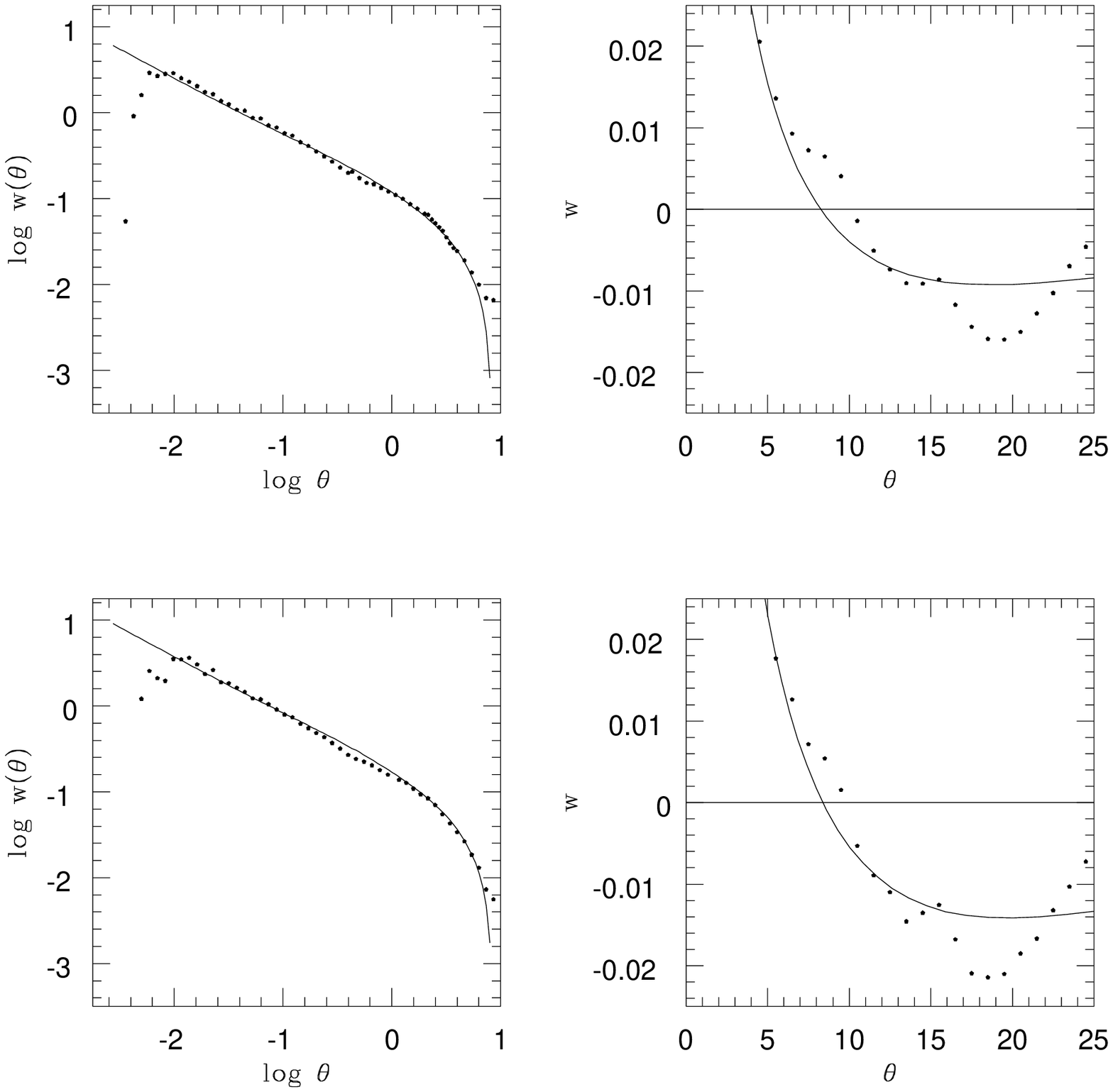}
\caption{ \small The measured and and from the 
Limber--equation, using selection function 2 and in (\ref{eq:xipar}) 
parametrised 3d CF calculated, 2d CFs for m=16.5 mag (upper panel),
m=16 mag (lower panel)} 
\end{figure}
\end{center}

\newpage
\begin{center}
\begin{figure}
\leavevmode \epsfxsize=5in \epsfbox{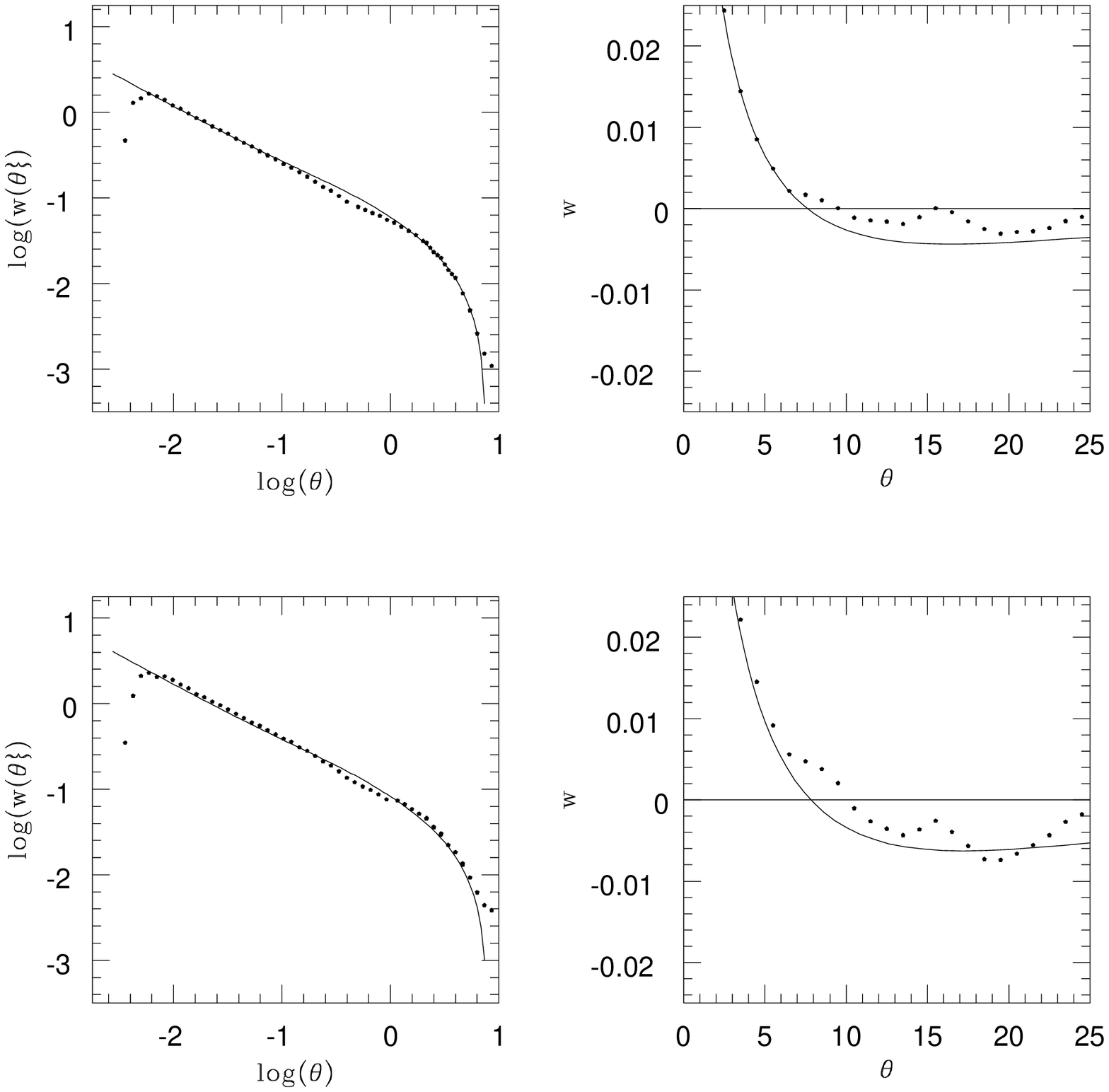}
\caption{ \small  The measured and and from the 
Limber--equation, using selection function 2 and in (\ref{eq:xipar}) 
parametrised 3d CF calculated, 2d CFs for m=17.5 mag (upper panel),
m=17 mag (lower panel)} 
\end{figure}
\end{center}

\newpage
\begin{center}
\begin{figure}
\leavevmode \epsfxsize=5in \epsfbox{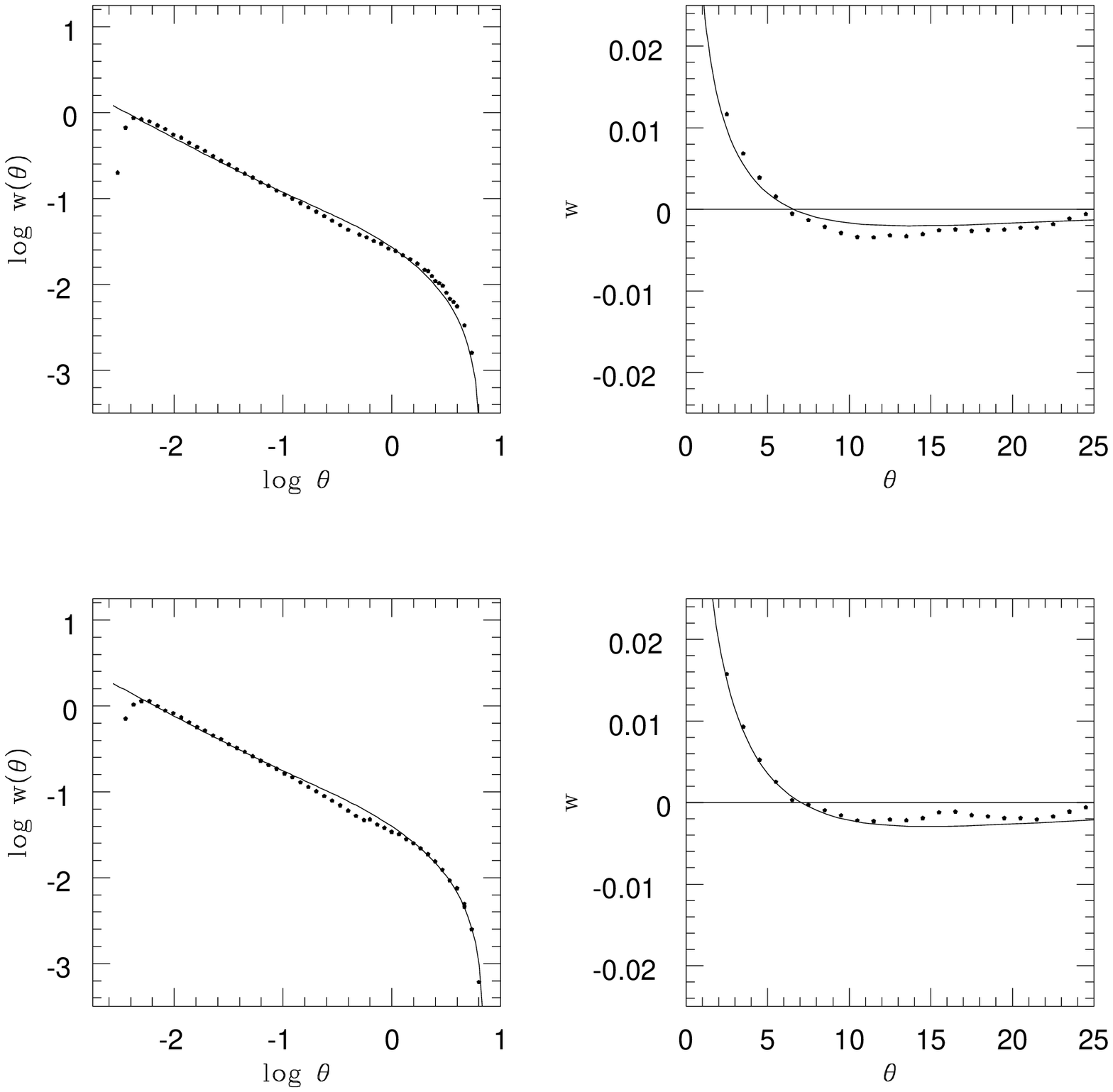}
\caption{ \small The measured and and from the 
Limber--equation, using selection function 2 and in (\ref{eq:xipar}) 
parametrised 3d CF calculated, 2d CFs for m=18.5 mag (upper panel),
m=18 mag (lower panel)} 
\end{figure}
\end{center}

\end{document}